# Bottom up synthesis of multifunctional nanoporous graphene

**Authors:** César Moreno,[1,*] Manuel Vilas-Varela,[2]† Bernhard Kretz,[3]† Aran Garcia-Lekue,[3,4] Marius V. Costache,[1] Marcos Paradinas,[1] Mirco Panighel,[1] Gustavo Ceballos,[1] Sergio O. Valenzuela,[1,5] Diego Peña,[2,*] and Aitor Mugarza[1,5,*]

**Affiliations:**

[1]Catalan Institute of Nanoscience and Nanotechnology (ICN2), CSIC and The Barcelona Institute of Science and Technology, Campus UAB, Bellaterra, 08193 Barcelona, Spain.

[2]Centro de Investigación en Química Biolóxica e Materiais Moleculares (CIQUS) and Departamento de Química Orgánica, Universidade de Santiago de Compostela. Santiago de Compostela 15782, Spain.

[3]Donostia International Physics Center, Paseo M. de Lardizabal 4, 20018 San Sebastian, Spain.

[4]Ikerbasque, Basque Foundation for Science, 48013 Bilbao, Spain

[5]ICREA Institució Catalana de Recerca i Estudis Avançats, Lluis Companys 23, 08010 Barcelona, Spain

*Correspondence to: cesar.moreno@icn2.cat (C.M.); diego.pena@usc.es (D. P.); aitor.mugarza@icn2.cat (A. M.)

†These authors contributed equally to this work.

**Abstract**: Nanosize pores can turn semimetallic graphene into a semiconductor and from being impermeable into the most efficient molecular sieve membrane. However, scaling the pores down to the nanometer, while fulfilling the tight structural constraints imposed by applications, represents an enormous challenge for present top-down strategies. Here we report a bottom-up method to synthesize nanoporous graphene comprising an ordered array of pores separated by ribbons, which can be tuned down to the one nanometer range. The size, density, morphology and chemical composition of the pores are defined with atomic precision by the design of the molecular precursors. Our measurements further reveal a highly anisotropic electronic structure, where orthogonal one-dimensional electronic bands with an energy gap of ~1 eV coexist with confined pore states, making the nanoporous graphene a highly versatile semiconductor for simultaneous sieving and electrical sensing of molecular species.

**One Sentence Summary:** We synthesize atomically-precise nanoporous graphene that combines transistor with molecular sensing and sieving functionalities.

**The main text:**

Nanoporous graphene has recently attracted great attention due to its potential application as an active component of field effect transistors (FET) (*1*, *2*) and as an atom-thick selective nanosieve for sequencing (*3*, *4*), ion transport (*5*, *6*), gas separation (*7–9*), and water purification (*10*, *11*). Selectivity in molecular sieving is achieved by reducing the pore size to the scale of single molecules, that is, in the nanometer range for relevant greenhouse gases, aminoacids, or single ions. This has been achieved in several studies at the single pore level (*6*) or by the creation of randomly distributed pores (*9*, *11*), where graphene remains semimetallic. Similarly, inducing semiconducting gaps for room temperature gate actuation requires the generation of sub-10 nanometer ribbons between pores (*1*, *12*). In this range, atomic scale disorder and width fluctuations have substantial impact on gap uniformity. Hence, combining semiconducting and sieving functionalities in a single nanoporous graphene material is a challenging task that requires the simultaneous generation of nanometer size pores and ribbons that have to be carved with atomic precision.

Inspired by successful on-surface routes to synthesize covalent carbon-based nanostructures (*13–20*), we have devised a strategy that leads to the formation of nanoporous graphene that exhibits both semiconducting and nanosieving functionalities. Our method relies on the hierarchical control of three thermally activated reaction steps, labelled as T1-T3 in the sketch of Fig. 1. Nanoribbons and pores with nanometer size, atomic-scale uniformity and long-range order are formed in separate steps. Graphene nanoribbons are first synthesized by following a previously used route (*17*, *20*), consisting on the surface-assisted Ullmann coupling of aromatic dihalide monomers into polymer chains (T1), and the cyclodehydrogenative aromatization of the intermediate polymeric chains into GNRs (T2). The final step (T3) interconnects GNRs laterally in a reproducible manner via a highly selective dehydrogenative cross coupling(*21*). This step requires a careful design of the monomer precursor, which defines the edge topology of the resulting GNR that is necessary for a high yield and selectivity of the cross coupling reaction. The monomer precursor synthesized in this work, labelled as DP-DBBA, is a derivative of the 10,10'-dibromo-9,9'-bianthracene, used in the synthesis of 7 carbon atom wide armchair GNRs (7-AGNR) (*17*), with phenyl substituents added at (2,2') sites. The latter is the key element for the promotion of the inter-GNR connections that lead to the nanoporous graphene (NPG) structure sketched in Fig. 1D (see Supplementary Materials for details in the monomer synthesis) (*22*). The choice of catalytic surface is also relevant for the selection of the reaction paths that define the intermediates and for the separation of thermal windows that lead to their hierarchical control. Here we use the Au(111) surface, where each reaction step has a different thermal activation onset, as shown below. The NPG can then be transferred to suitable substrates in which their functionalities can be exploited (*22*).

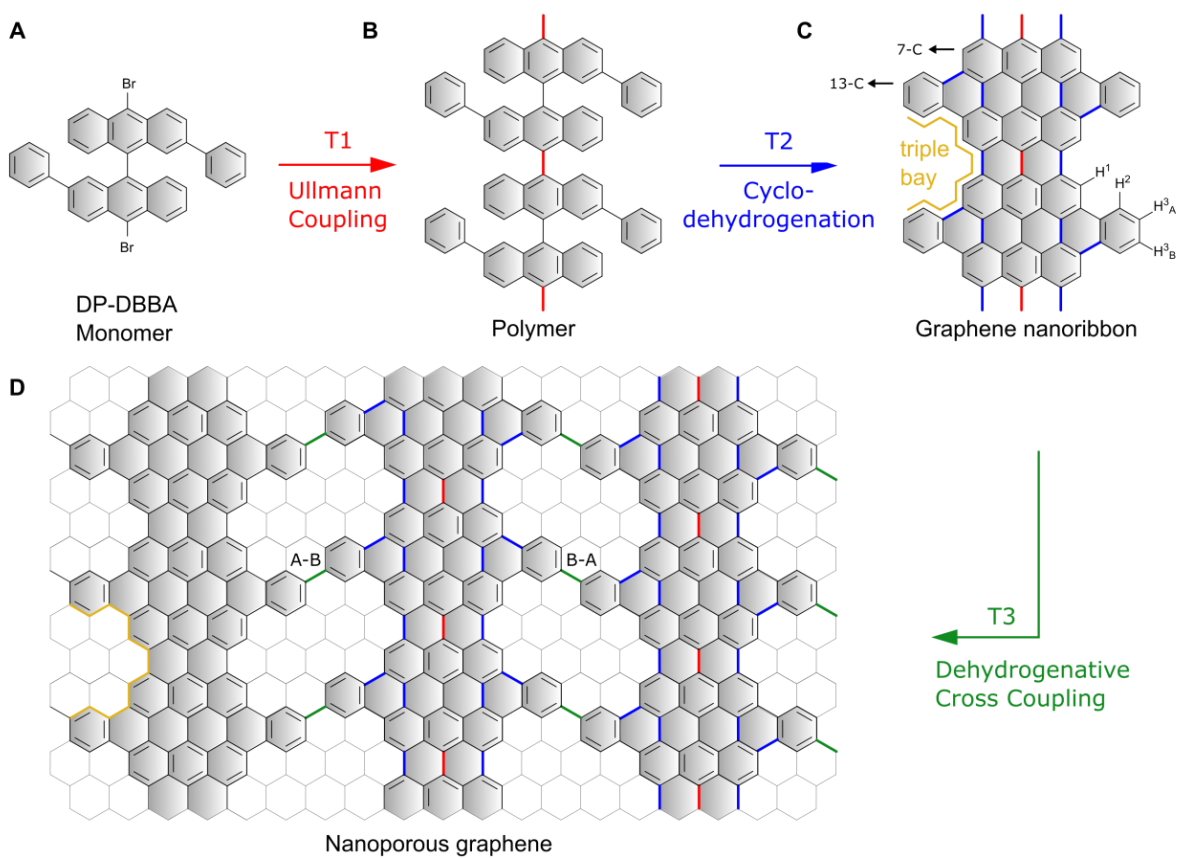

**Fig. 1**. **Schematic illustration of the synthetic hierarchical path for the generation of nanoporous graphene.** (**A**) The DP-DBBA monomer used as precursor. (**B**) At T1, DP-DBBA is debrominated and the radical carbon atoms cross couple to form polymer chains. (**C**) At T2, an intramolecular cyclodehydrogenation leads to the planar graphene nanoribbon. The cyclization of the phenyl substituent modulates the width of the GNR with pairs of 7 and 13 C atom wide sections, forming multibay regions that consist of three conjoined bay regions (yellow lines), and leaving three types of C-H bonds at the edge ($H^{1-3}$). Each type will have two equivalent positions, as represented for $H^3$ with A and B labels. (**D**) Finally, at T3 the GNRs are interconnected from the $H^3$ bonds via dehydrogenative cross coupling, giving rise to the nanoporous graphene structure (the extended graphene structure is underlaid to highlight the structure of the nanopores). The A-B or B-A bonding combinations give rise to identical pores with different orientations.

The structures obtained in each step of the hierarchical synthetic route are characterized using scanning tunneling microscopy (STM). Representative topographic images are shown in Fig. 2. After deposition at room temperature and annealing to $T = 200°C$, monomers undergo debromination to form the corresponding aryl radicals which are subsequently coupled by means of C-C bond formation (T1) (*17*, *20*). The resulting polymeric chains exhibit the characteristic protrusion pairs with a periodicity of 0.84 nm and an apparent height of 0.31 nm, which arises from probing the high-ends of the staggered bis-anthracene units of the monomer with STM (Fig. 2, A and D) (*17*, *20*). The chains, with lengths of up to 150 nm, predominantly align in close-packed ensembles along the zig-zag orientation of the herringbone reconstruction of the Au(111) surface. Both the extraordinary length of the polymeric intermediates and their parallel alignment are crucial ingredients for the high yield and long-range order observed in the final step T3.

Annealing to $T = 400°C$ triggers the intramolecular cyclodehydrogenation (T2), giving rise to the aromatization of the chain and the corresponding reduction of the apparent height to $h = 0.18$ nm, which is characteristic of GNRs (Fig. 2, B and E) (*17*, *20*). The nanoribbons appear dispersed as individual stripes, yet they maintain a predominantly parallel alignment along the zig-zag orientation. As can be seen in the high-resolution image of Fig. 2E, the cata-fused benzene rings that arise from the cyclization of the phenyl substituent, result in a periodic modulation of the width. Consecutive pairs of 7 and 13 C atoms define multibay regions made of three conjoined bays (yellow lines in Fig. 1, C and D). This particular edge structure of the nanoribbons, labelled as 7/13-AGNR hereafter, will define both the morphology and size of the corresponding pores in the NPG and its electronic structure.

The aryl-aryl inter-ribbon connection is induced by further annealing to $T = 450°C$ (T3). Figure 2C shows how GNRs tend to merge, connecting laterally from each of the fused benzene rings and forming a nanomesh (green square). The submolecular structure, observed in the high-resolution image of Fig. 2F, coincides with the NPG structure depicted in Fig. 1D, which reveals that the inter-ribbon coupling occurs via a selective C-H[3] bond activation. The activation of specific C-H bonds in polycyclic aromatic hydrocarbons is non-trivial due to the presence of multiple quasienergetic bonds (three in the case of the 7/13 AGNR, labelled as H$^{1-3}$ in Fig. 1). In step T3, the selectivity in the C-C bond formation between GNR is driven by the easy accessibility to the radical formed after the C-H$^3$ bond cleavage, as opposed to the steric hindrance associated with the radicals formed after the C-H$^1$ or C-H$^2$ bond cleavage. Another remarkable milestone is the long-range order achieved. To date, the observation of selective intermolecular aryl-aryl coupling has been limited to small supramolecular structures (*19*, *23–28*). The hierarchical strategy of our method allows us to set the long-range order in step T2, where the length of prealigned GNRs represent the size limitation of the NPG. The high yield and remarkable selectivity of this coupling mechanism can be best appreciated when the surface is saturated with polymeric chains by depositing the precursor with the substrate at *T* = 200°C (Fig. 2, G and H). As a result, a coupling yield close to 100% is achieved, where every GNR is integrated in a NPG domain. Following this procedure NPG sheets as large as 70 × 50 nm$^2$ are easily obtained, with atomically reproducible pores of 0.9 × 0.4 nm$^2$, ultra-high densities of 480 × 10$^3$ pore/μm$^2$, and a characteristic defect concentration of ~2%.

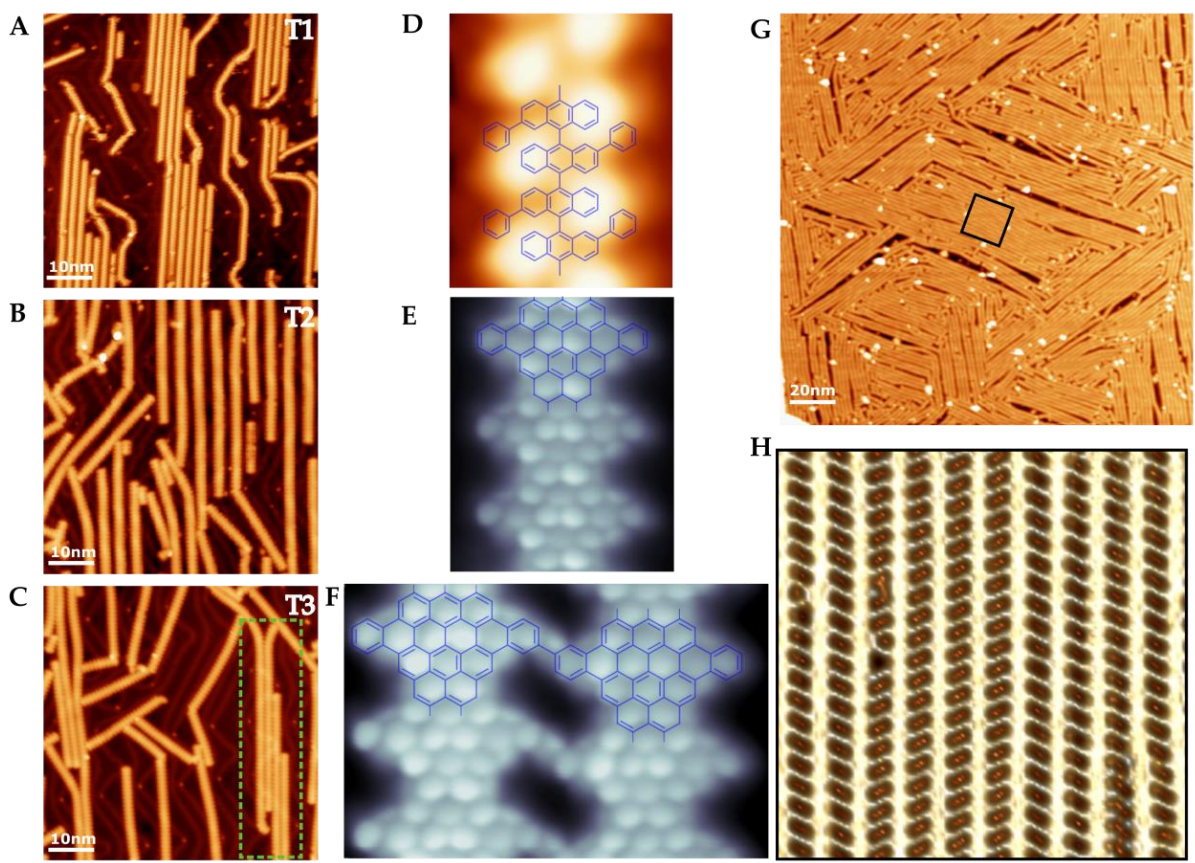

**Fig. 2. Hierarchical synthesis of nanoporous graphene.** (**A-C**) Constant current STM images showing the distribution and morphology of the different covalent structures obtained for a low coverage of precursors after each thermal annealing step T1, T2 and T3. At this coverage, NPG stripes form locally (green square in C). (**D-F**) Zoomed images revealing the internal structure in each case. The high resolution images in **E** and **F** are obtained by using a CO-functionalized tip in constant height mode (*22*). The atomic models depicted in Fig. 1 for T1-T3 are overlaid in D-F respectively. (**G**) Constant current STM image of a surface totally covered with NPG domains with sizes up to 70 × 50 nm$^2$, obtained by a saturated deposition of the precursor at T = 200°C. (**H**) Laplacian filtered topographic close-up image of the NPG region marked in **G**, showing a regular array of identical pores with low defect density. All imaging parameters in Supplementary Material (*22*).

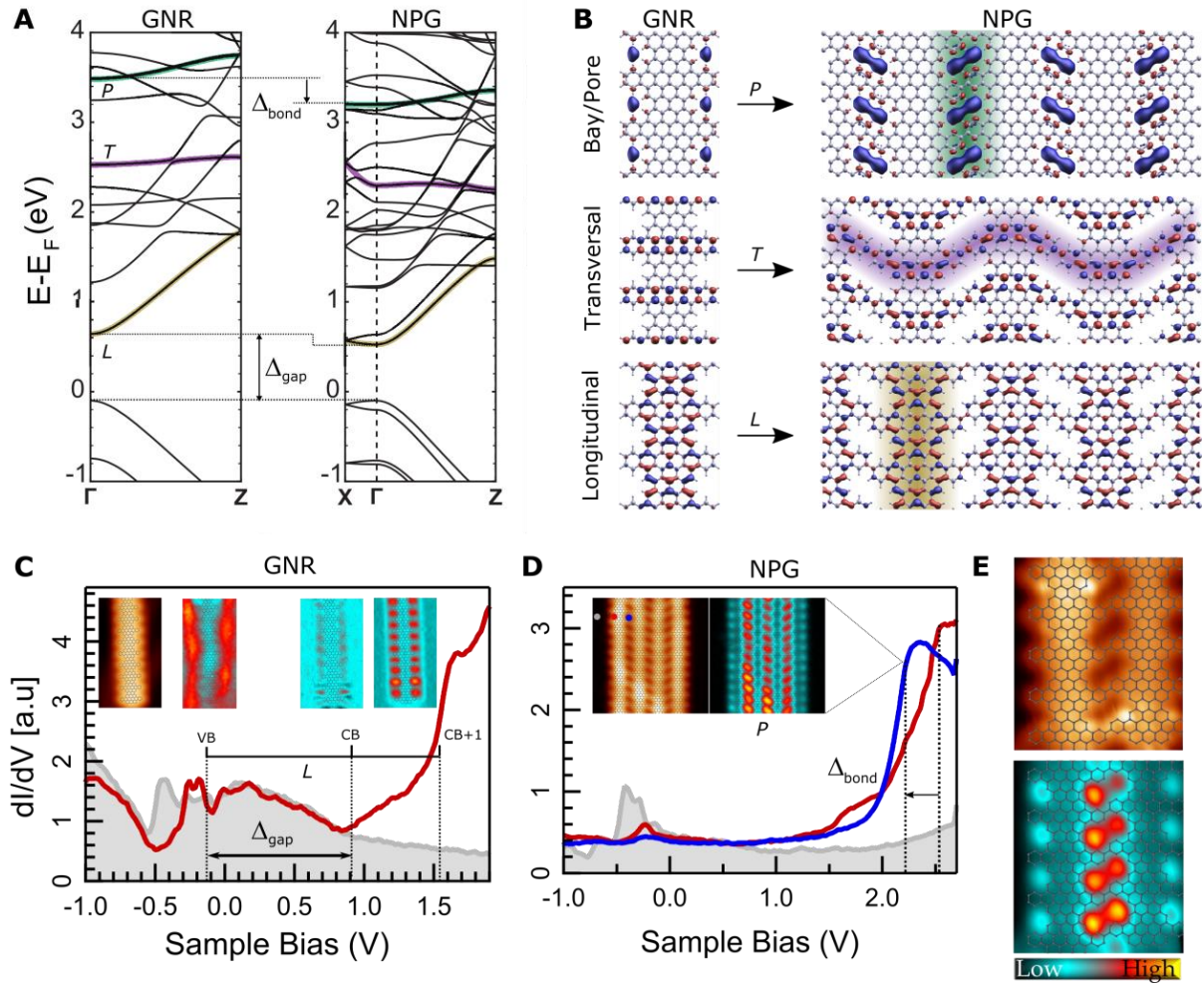

**Fig. 3. Electronic properties of the 7/13 AGNR and the NPG.** (**A**) Band structure calculated by DFT for individual 7/13-AGNRs (left) and the NPG (right). Examples of longitudinal (*L*), transversal (*T*) and bay/pore (*P*) bands are highlighted with yellow, purple and green guiding lines respectively. The Fermi level is determined by using the experimental VB energy as reference (*22*). (**B**) Wave functions at Γ for each of the band examples in **A**. Their dispersion direction is highlighted by guiding stripes. (**C**) dI/dV spectra acquired at the multibay edge of a 7/13-AGNR, where the onset of the CB, VB and CB+1 bands can be identified. (**D**) dI/dV spectra acquired at the peripheral multibay (red) and a pore (blue) region of an NPG. The interaction between the two bay states within a pore results in an energy shift of $\Delta_{bond}$ due to the formation of a bonding band. In **C** and **D** reference spectra acquired on Au(111) is added in shaded gray, and the insets show constant height tunneling current ($I_t$) images (left) and dI/dV maps (right) acquired at the related energies (**E**) High resolution constant height $I_t$ image (top) and dI/dV map (bottom) of the pore states, acquired at +2.2 eV, where the double lobe structure predicted by DFT is reproduced. The localization of this state within the pore is shown by overlaying an atomic model of the local NPG structure.

The peculiar topology of the NPG imprints a band gap, 1D anisotropy and different type of localization in the electronic states, with potential implications in transport and sensing. These can be rationalized within the same hierarchical approach used in the synthesis, namely by considering the states of individual 7/13-AGNR as building blocks that already contain the main features, and following their evolution as the ribbons interconnect. By combining density functional theory (DFT) (Fig. 3, A and B) and scanning tunneling spectroscopy (STS) (Fig. 3, C-E), three types of bands are identified. Examples of each band, and its corresponding wave function, are highlighted in different colors in Fig. 3, A and B, respectively: *Longitudinal bands* (*L*, yellow), *Transversal bands* (*T*, purple), and *Bay and Pore bands* (*P*, green).

*L* and *T* bands originate from the carbon *s* and *p* orbitals. The *L* bands are similar to the conventional bands in straight armchair GNRs and disperse along the ribbon (along ΓZ). They appear confined within the 7-C atom wide backbone of the GNR. On the contrary, the *T* bands are localized within the 13-C atom wide periodic stripes, and thus they do not disperse in the longitudinal direction. They arise from the superlattice periodicity imprinted by the modulated width of the 7/13-AGNR and are therefore exclusively related to its edge topology. The semiconducting gap of the 7/13-AGNR is determined by the *L* bands. For the free-standing ribbon, density functional theory (DFT) predicts a bandgap of 0.74 eV, which is increased to 1.36 eV after including self-energy corrections within the GW approximation (*22*). Experimental STS spectra reveals a bandgap of $\Delta_{gap}$ = 1.0 eV (Fig. 3C), slightly lower than the GW bandgap, as expected from the screening effect of the underlying substrate (*29*, *30*). Notably, this value is smaller than the 1.5 eV measured for the wider 13-AGNR (*29*), highlighting the role of edge topology on determining band gaps.

The effect of the inter-ribbon connection is specific to the band type (Fig. 3A). Protected within the backbone, *L*-bands remain unperturbed in the NPG, as indicated by the lack of dispersion in the transversal direction (along ΓX). The DFT bandgap is only reduced by 0.12 eV when compared with GNRs, which agrees with a downshift of similar size measured by STS for the conduction band onset (*22*). In contrast, the extension of the *T*-band wave functions across the 13-C atom wide section enables substantial inter-ribbon coupling and the formation of 1D dispersing states with a similar mobility as the longitudinal ones. The resulting wave functions consist of non-interacting zigzag stripes that run across the GNRs. In the shown calculations, the structure alternates between A-B and B-A pores, but the same conclusions are obtained using NPGs formed exclusively by either of the two configurations (*22*).

The origin of the *P* bands is more exotic. Localized within the vacuum pocket defined by the multibay region, they are not related to atomic orbitals or their hybridization, as has been observed in other molecular pores (*31*). Instead, they originate from the free-electron like image potential states that are confined at the vacuum side along the GNR edge. They can be regarded as the 2D analogue of the superatom states that develop when a graphene sheet bends into a fullerene (*32*). In our lower dimension analogue, the straight graphene edge "bends" into a periodic array of weakly coupled multibays that give rise to rather flat superatom bands (*22*). In the NPG, the *P* states of adjacent multibays interact when they pair to form pores, leading to bonding and antibonding bands, as observed in the DFT band structure (*22*). Experimentally, the high-energy lying antibonding band cannot be probed without affecting the integrity of the NPG. STS spectra, however, does reveal an energy reduction of $\Delta_{bond}$ = 0.30 eV, which corresponds to the formation of the bonding band (Fig. 3D) and is in very close agreement with the shift of 0.28 eV obtained by DFT.

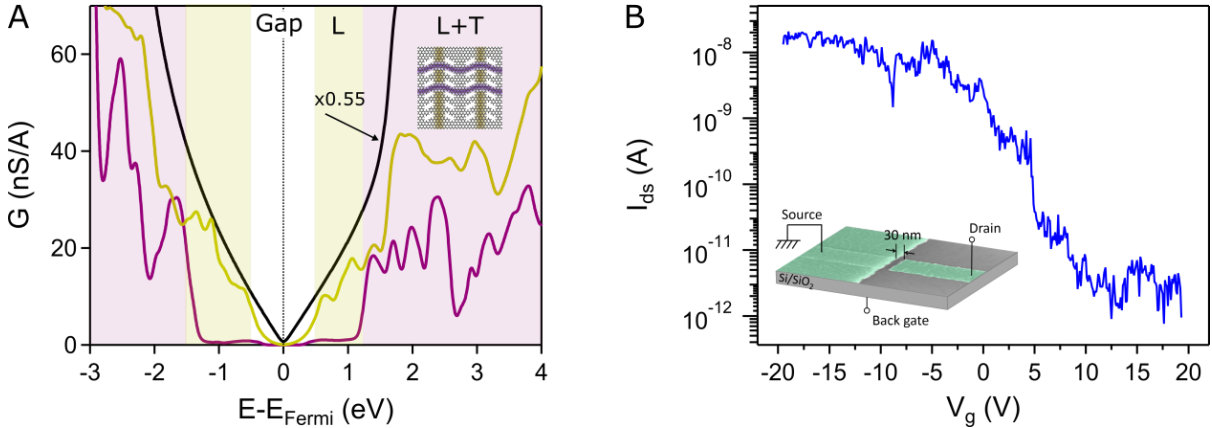

**Fig. 4. Transport properties of the NPG.** (**A**) Conductance calculated in the longitudinal (yellow) and transversal (purple) directions of the NPG as defined in Fig. 3, with that of pristine graphene added as comparison (black line). Colored backgrounds delimit regions of total gap (white), only longitudinal (yellow) and combined longitudinal and transversal (purple) transport. (**B**) $I_{ds}$−$V_g$ characteristics of a NPG device of 30 nm channel length, gated by a 90 nm thick $SiO_2$ gate oxide.

The band heterogeneity described above is reflected in the energy-dependent conductance across the two perpendicular directions (Fig. 4A). Three regions can be identified in the calculations, depending on the energy $E$. A region of true energy gap around the Fermi energy $E_F$, in which the conductance is fully suppressed, a region for $|E-E_F| < 1.2$ eV, in which transport is purely longitudinal, and a region for $|E-E_F| > 1.2$ eV, in which transport has both longitudinal and transversal components. In order to experimentally demonstrate the semiconducting properties of the nanoporous graphene, which is expected about the bandgap, the transport response was characterized using field-effect transistor structures. The NPG was first transferred onto a $Si/SiO_2$ substrate and then contacted with Pd electrodes using electron-beam lithography and shadow evaporation (*22*). The Si substrate is highly-doped to fulfil the role of a backgate electrode across the 90-nm thick $SiO_2$ gate dielectric. Remarkably, the large dimension of the NPG sheets enables a large device yield of ~75% for the designed channel length of 30 nm. Figure 4B shows typical room-temperature current-gate ($I_{ds} - V_g$) behavior at fixed drain-source voltage bias ($V_{ds}$). The devices show good performance, presenting hole-transport and an on-off ratio of ~$10^4$, which is comparable to prior work on single GNRs (*33*). The transistor characteristics are highly non-linear at low bias, indicating that the transport is limited by the presence of a Schottky barrier at the Pd-NPG interface (*21,33*) and suggesting that larger on-off ratios could be obtained by lowering the contact work-function (*33*).

The semiconducting functionality of the presented NPG architecture can be exploited in a new generation of graphene-based devices such as FET-sensors or gate-controlled sieves. At high doping level, the onset of the transversal bands provides an orthogonal, non-interacting 1D channel. The set of **L** and **T** bands brings to graphene the in-plane anisotropy that makes 2D materials, such as phosphorene or black phosphorous, appealing for FET, optical and sensing applications (*34, 35*). Finally, the presence of confined states within the nanopores makes NPG very attractive for detection and electronic tracking in chemical and bio sensors and filters. Confined states could be shifted down to the Fermi level by the interaction with ions and molecules (*36, 37*), making them detectable in transport measurements.

**Acknowledgments:** We acknowledge thoughtful discussions with Enrique Guitián and Dolores Pérez for the synthesis of the monomer precursors. We thank Pedro Brandimarte and Daniel Sanchez-Portal for useful discussions related to the calculations. We thank E. Del Corro, C. J. J. Hebert and J. A. Garrido for the support with the Raman measurements. **Funding:** This research was funded by the CERCA Programme/Generalitat de Catalunya and supported by the Spanish Ministry of Economy and Competitiveness, MINECO (under Contract No. MAT2016-78293-C6-2-R, MAT2016-78293-C6-3-R, MAT2016-78293-C6-4-R, MAT2016-75952-R and Severo Ochoa No. SEV-2013-0295), the Secretariat for Universities and Research, Knowledge Department of the Generalitat de Catalunya 2014 SGR 715, 2014 SGR 56 and 2017 SGR 827, the Basque Department of Education (Contract No. PI-2016-1-0027), the Xunta de Galicia (Centro singular de investigacion de Galicia accreditation 2016–2019, ED431G/09), and the European Regional Development Fund (ERDF) and the European Union's Horizon 2020 research and innovation programme under Grant Agreement 696656. C.M was supported by the Agency for Management of University and Research grants (AGAUR) of the Catalan government through the FP7 framework program of the European Commission under Marie Curie COFUND action 600385. M. Par. thanks the Spanish Government for financial support through PTA2014-09788-I fellowship. **Author contributions:** D. P. and M. V.-V. synthesized the monomer precursor, B. K. and A. G.-L. performed the ab-initio calculations. C. M. and M. Pan. performed the STM measurements. C. M., G. C. and A. M. analyzed the STM data. M. Par. and C. M. carried out the sample transfer and Raman characterization. M. V. C. and S. O. V. fabricated the devices and carried out the transport measurements. All authors discussed the results and participated in writing the manuscript. D. P., G.C. and A. M. initiated and directed this research. **Competing interests:** C.M., A.M., G.C., D.P. and M.V.-V. have been filed a EU pending patent. **Data and materials availability:** All data needed to evaluate the conclusions in the paper are present in the paper or the Supplementary Materials.

**Supplementary Materials:**
Materials and Methods
Supplementary Text
Figs. S1 to S12